\documentclass[prd,twocolumn,showpacs,reprint,preprintnumbers,nofootinbib,superscriptaddress]{revtex4-1}

\input{header}
\usepackage{fontawesome5}
\usepackage{cancel}

\newcommand{\bea}{\begin{eqnarray}\begin{aligned}}
\newcommand{\eea}{\end{aligned}\end{eqnarray}}
\newcommand{\gd}{g_d}

\newcommand{\eq}{\text{eq}}

\newcommand{\MeV}{\text{MeV}}
\newcommand{\GeV}{\text{GeV}}

\newcommand{\fo}{\text{FO}}
\newcommand{\fii}{\text{FI}}

\newcommand{\eff}{\text{eff}}

\newcommand{\rh}{\text{rh}}

\newcommand{\mpl}{m_\text{pl}}

\newcommand{\rad}{\text{rad}}

\newcommand{\alphaem}{\alpha_\text{em}}

\newcommand{\Eq}[1]{Eq.~(\ref{eq:#1})}

\newcommand{\Sec}[1]{Sec.~\ref{sec:#1}}
\newcommand{\Subsec}[1]{Sec.~\ref{subsec:#1}}
\newcommand{\Appx}[1]{Appendix~\ref{appx:#1}}

\newcommand{\Fig}[1]{Fig.~\ref{fig:#1}}

\newcommand{\dth}{\text{dth}}

\newcommand{\chibar}{\bar{\chi}}
\newcommand{\fbar}{\bar{f}}

\newcommand{\DM}{{\rm DM}}
\newcommand{\SM}{{\rm SM}}
\newcommand{\QCD}{\text{QCD}}

\def\approxprop{%
  \def\p{%
    \setbox0=\vbox{\hbox{$\propto$}}%
    \ht0=0.6ex \box0 }%
  \def\s{%
    \vbox{\hbox{$\sim$}}%
  }%
  \mathrel{\raisebox{0.7ex}{%
      \mbox{$\underset{\s}{\p}$}%
    }}%
}

\begin{document}

\preprint{UCI-HEP-TR-2023-05; FERMILAB-PUB-23-428-T-V}

\title{Cosmic Millicharge Background and Reheating Probes}

\author{Xucheng Gan}
\email{xg767@nyu.edu}

\affiliation{Center for Cosmology and Particle Physics, Department of Physics, New York University, New York, NY 10003, USA}

\author{Yu-Dai Tsai}
\email{yudait1@uci.edu}
\email{yt444@cornell.edu}
\affiliation{Department of Physics and Astronomy,  University of California, Irvine, CA 92697-4575, USA}
\affiliation{Fermi National Accelerator Laboratory (Fermilab), Batavia, IL 60510, USA}
\begin{abstract}

We demonstrate that the searches for dark sector particles can provide probes of reheating scenarios, focusing on the cosmic millicharge background produced in the early universe. We discuss two types of millicharge particles (mCPs): either with, or without, an accompanying dark photon. These two types of mCPs have distinct theoretical motivations and cosmological signatures. We discuss constraints from the overproduction and mCP-baryon interactions of the mCP without an accompanying dark photon, with different reheating temperatures. We also consider the $\Delta N_{\rm eff}$ constraints on the mCPs from kinetic mixing, varying the reheating temperature. The regions of interest in which the accelerator and other experiments can probe the reheating scenarios are identified in this paper for both scenarios. These probes can potentially allow us to set an upper bound on the reheating temperature down to $\sim 10$ MeV, much lower than the previously considered upper bound from inflationary cosmology at around $\sim 10^{16}$ GeV. In addition, we find parameter regions in which the two mCP scenarios may be differentiated by cosmological considerations. Finally, we discuss the implications of dedicated mCP searches and future CMB-S4 observations.

\end{abstract}

\maketitle

\tableofcontents

\section{Introduction}
\label{sec:intro}

The study of millicharged particles (mCPs), meaning particles with small rational or irrational electric charges, is closely connected to the studies of charge quantization~\cite{Dirac:1931kp,Schwinger:1966nj, Zwanziger:1968rs}, Grand Unification Theories (GUTs)~\cite{Pati:1973uk,Georgi:1974my,Preskill:1984gd}, string theory~\cite{Wen:1985qj} and compactifications \cite{Burgess:2008ri, Goodsell:2009xc,Cicoli:2011yh,Shiu:2013wxa,Feng:2014eja}. 
More on the phenomenological and cosmological side, dark matter (DM) \cite{Brahm:1989jh, Feng:2009mn, Cline:2012is, Chu:2011be}, dark sector \cite{Okun:1982xi,Holdom:1985ag}, and $21\text{-cm}$ cosmology~\cite{Munoz:2018pzp, Berlin:2018sjs, Barkana:2018qrx, Slatyer:2018aqg, Kovetz:2018zan,  Creque-Sarbinowski:2019mcm,Liu:2019knx,Mathur:2021gej}, further stimulate the mCP studies.
These motivations incentive searches of mCP and millicharged dark matter (mDM) at all fronts, including fixed-target experiments~\cite{Golowich1987,Babu:1993yh,Prinz:1998ua,Badertscher:2006fm, Bjorken:2009mm, NA62:2017rwk,Magill:2018tbb, Berlin:2018bsc,SHiP:2018yqc,Kelly:2018brz,Harnik:2019zee,ArgoNeuT:2019ckq,Marocco:2020dqu}, colliders~\cite{Davidson:2000hf,CMS:2012xi,Jaeckel:2012yz,Haas:2014dda,Ball:2016zrp,Alvis:2018yte,Liu:2018jdi, Chu:2018qrm,Chu:2020ysb,milliQan:2021lne}, reactors~\cite{Gninenko:2006fi,TEXONO:2018nir},
large neutrino observatories~\cite{Hu:2016xas, Arguelles:2019xgp,Plestid:2020kdm}, quantum sensors~\cite{Afek:2020lek, Budker:2021quh, Yu:2021qde}, cavity experiments~\cite{Gies:2006hv,Ahlers:2006iz,Silva-Feaver:2016qhh, Berlin:2019uco, Berlin:2021kcm,Romanenko:2023irv, Berlin:2023gvx}, and direct-detection experiments~\cite{Erickcek:2007jv,Agnese:2014vxh,Hambye:2018dpi, Mahdawi:2018euy, Emken:2019tni,Berlin:2021zbv,Li:2022idr,SENSEI:2023gie,Oscura:2023qch}. The imprints on cosmic microwave background (CMB)~\cite{Dubovsky:2003yn, Dolgov:2013una,  dePutter:2018xte, Xu:2018efh, Kovetz:2018zan, Buen-Abad:2021mvc, Berlin:2022hmt, Vogel:2013raa, Adshead:2022ovo} and astrophysical observations~\cite{Dobroliubov:1989mr,Davidson:2000hf, Vogel:2013raa, Vinyoles:2015khy, Chang:2018rso, Wadekar:2019mpc, Chu:2019rok, Stebbins:2019xjr, Berlin:2021kcm,  Lasenby:2020rlf, Cruz:2022otv, Chang:2022gcs} are also explored. 
The studies of mCP and mDM have taken the critical stage of the Beyond the Standard Model (BSM) research in particle physics, astroparticle physics, and cosmology.

The reheating process after the inflation has significant impacts on the cosmic evolutions and the late-time abundances of dark-sector particles. However, there has not been a solid determination of the reheating temperature. 
In this paper, we reverse the logic and demonstrate that the searches for dark sector particles can provide tests for the reheating scenarios, using mCP as an example.
We consider probes that can set an upper bound on the reheating temperature down to $\sim 10\,\MeV$, much lower than the previously considered upper bounds from inflation cosmology~(for example, the tensor-to-scalar ratio constrains the energy density at the end of inflation, which gives an upper bound around $\sim 10^{16}\,\GeV$~\cite{Lyth:1984yz, Lyth:1987aa, Lyth:1996im, Planck:2015sxf,Domcke:2015iaa}).

mCPs, without the assumption of their initial abundance before reheating, can be irreducibly generated through the reheating process. We thus refer to it as cosmic millicharged background (CmB). CmB abundance would depend on the reheating scenarios and temperatures, which we will explore further.
In this paper, we provide strong motivations and novel views for mCPs searches, focusing on the theoretical and cosmological aspects.
The main focus is to study the CmB cosmology {\it with or without} an accompanying dark photon.

mCPs are produced through the freeze-in process from the SM bath during reheating. mCP without a massless dark photon can easily be overproduced without the annihilation channel to massless dark photons. The produced mCPs can be constrained by the measurements of current DM abundance, as well as additional CMB constraints from mCP-baryon interactions.
However, these constraints are highly dependent on the reheating temperature. We carefully discuss the mCP overproduction bound from solving the Boltzmann equation, considering different reheating temperatures.

mCPs with an accompanying dark photon have an additional annihilation channel to deplete the number density, and can easily avoid the overproduction bound. However, 
the dark radiation composed of the massless dark photon, $A'$, will affect the
CMB and BBN $\Delta N_\eff$ measurements of the relativistic degrees of freedom. These measurements then provide us with constraints on the mCP with dark photon scenario, also highly depending on the reheating temperatures. Our new results include such dependence on the reheating temperatures.

In this paper, we meticulously consider these two types of constraints, and identify the regions of search in which we can test the reheating scenarios. 
The existing and upcoming dedicated mCP detectors, including milliQan (constructed and taking data) \cite{Ball:2016zrp, milliQan:2021lne}, FerMINI \cite{Kelly:2018brz}, SUBMET (fully approved) \cite{Choi:2020mbk, Kim:2021eix}, and FORMOSA (demonstrator under construction for early 2024 installation)~\cite{Foroughi-Abari:2020qar}, are the leading searches for particles with eV energy deposits and coincidence signatures in multiple scintillator bars with small photon yields. Without having a small electric charge or ultralight mediators (at least smaller than eV), it is challenging for other dark sector or any Standard Model (SM) particles to have such a signature (e.g., SM neutrinos rarely have eV energy deposit). In these experiments, muon would be vetoed due to large photon yield, and long-lived particles do not have coincidence multi-bar signatures, which also reduce the dark noises to a negligible level. Thus, these dedicated searches have some of the best opportunities to confirm the discovery of mCP and test reheating scenarios, compared to other searches that have more BSM targets with degenerate signatures and/or rely on DM assumptions \cite{Rich:1987st, Erickcek:2007jv, Emken:2019tni,Afek:2020lek}. 
 The analyses of different reheating scenarios and tests can be extended to other irreducible dark sector particles, including axions \cite{Langhoff:2022bij}.

In addition, we discuss the theoretical implications of these cosmological considerations. mCP with or without dark photons are thought to be difficult to differentiate, given that the massless dark photon is decoupled from the SM sector (unless there exists an ambient mCP background that induces the dark plasmon mass for the dark photon).
However, given the distinctive cosmology, one can identify regions in which one of these scenarios is favored. Recently, new searches were developed to study the dark photon coupling of the kinetic-mixing mCP relying on the effect that the dark plasmon mass from mCPs enhances the photon-dark photon oscillation~\cite{Berlin:2022hmt,Berlin:2023gvx}, which further motivate our cosmological study to identify the region which favors the kinetic-mixing mCP.

Furthermore, we attempt to clear a common misconception.
mCP with dark photon can potentially mimic the ``pure" mCP, even in cosmological consideration, by turning the dark coupling $g_d$ small, making the channels between mCP $\chi$ and $A'$ ineffective. 
However, one cannot naively assume that mCP with a massless dark photon can mimic ``pure" mCP with arbitrary charges. Given the requirement of positive definite of the kinetic-term matrix, the kinetic mixing term $\epsilon FF'$ has a requirement that $\epsilon<1$~\cite{Weinberg:1996kr, Burgess:2008ri}. So there is a region in which mCP with a dark photon cannot completely mimic the ``pure" mCP, allowing us to differentiate the two scenarios, as pointed out in this paper.

The paper is organized as follows. In \Sec{two_mCP}, we introduce the two major mCP scenarios considered in this paper. In \Sec{cosmology}, we discuss the CmB of both ``pure" and kinetic-mixing mCPs and their cosmological implications.
We discuss the regions in which dedicated searches can test reheating scenarios, and how to differentiate the two types of mCPs, in \Sec{results}.
The detailed derivations of our results, including the calculations of the Boltzmann equation, can be found in the Appendix.

\section{Two mCP scenarios}
\label{sec:two_mCP}

We consider two major mCP scenarios. In both cases, mCPs 
have small hypercharges under SM $U(1)_Y$, but they have different origins as described below.
In this paper, we consider mCPs to be fermionic, but our analyses can be extended to scalar mCPs with only quantitative changes.

\vspace{-2mm}

\subsection{mCP from kinetic mixing}

Here we consider a particle $\chi$ charged under an additional dark gauge symmetry $U(1)_d$ with the gauge field $A'$, coupling $g_d$, and $\alpha_d \equiv g_d^2/4 \pi$.
The gauge field (named dark photon) $A'$ has a kinetic mixing with SM hypercharge $U(1)_Y$ gauge field $B$. The model is described by the following Lagrangian
\bea
\label{eq:mcp_KM_L}
\mathcal{L} & \supset i \chibar(\slashed{\partial} - i \gd \slashed{A}' + m_\chi) \chi \\
& \quad - 
\frac{1}{4} B_{\mu \nu} B^{\mu \nu} - \frac{1}{4} A'_{\mu \nu}A'^{\mu \nu} + \frac{\epsilon}{2 \cos \theta_w} B_{\mu\nu} A'^{\mu \nu}.
\eea
Here, $A'_{\mu \nu} = \partial_\mu A'_\nu - \partial_\nu A'_\mu$ is the field strength of the dark photon, $B_{\mu \nu} = \partial_{\mu} B_\nu - \partial_\nu B_\mu$ is the field strength of the $U(1)_Y$ gauge field $B$, and $\epsilon$ is the kinetic mixing between $U(1)_Y$ and $U(1)_d$.

For the case of massless dark photon $A'$ and unbroken $U(1)_d$, one can find a convenient basis by diagonalizing the kinetic terms and removing the mixing (up to $O(\epsilon^2)$, assuming $\epsilon\ll 1$) with a field redefinition $A' \rightarrow A' + \frac{\epsilon}{\cos \theta_w} B$. 
In the new basis, the Lagrangian becomes 
\bea
\label{eq:chi_B_coupling}
    \mathcal{L} \supset g' q_{\chi}\chibar \gamma^\mu \chi  B_{\mu}, \,\, \text{where $q_\chi = \frac{\epsilon g_d}{e}$ }.
\eea
Here, $g'=e/\cos\theta_w$ is the gauge coupling of $U(1)_Y$.
After the electroweak symmetry breaking, $B = \cos \theta_w A - \sin \theta_w Z$, therefore, $\chi$ becomes an mCP with an effective electric charge $q_\chi$. 

For a given value of $q_\chi$, there is extra freedom to choose the value of $g_d$ (and thus determine the value of $\epsilon$). As we will discuss in the following sections, the choice of $g_d$ can affect the cosmic evolution of both $\chi$ and $A'$ in the early universe.
In this paper, for the mCP from kinetic mixing, we will consider 
sizable $\alpha_d$ (we will discuss the value in the next section).
Theoretically, the natural coupling strength that $\alpha_d \sim \alphaem$ can, for example, originate from the mirror symmetry between the SM sector and the dark sector~\cite{Chacko:2005pe,Chang:2006ra,Craig:2015pha,Dunsky:2019api, Batell:2019ptb,Liu:2019ixm, Batell:2022pzc}.

The value of $\epsilon$ can have large variations given the UV models. For example, one can consider $\epsilon$ stemming from the 1-loop contribution of the doubly charged heavy messengers~\cite{Holdom:1985ag}, i.e., 
\bea
\label{eq:1_loop_eps}
\epsilon \sim \frac{N Q_Y Q_d e g_d}{16 \pi^2}.
\eea
Here, $Q_Y$, $Q_d$, and $N$ are the $U(1)_Y$ charge, $U(1)_d$ charges, and the flavor number of the heavy messengers, respectively. 
If one chooses $Q_Y, Q_d, N \sim 1$, then $\epsilon \sim e g_d/16 \pi^2$.
In addition, larger or smaller $\epsilon$ can also be achieved. For large $\epsilon$, as long as it does not exceed $1$, it can be achieved by increasing the combination of $N Q_Y Q_d$.
For $\epsilon\ll e g_d/16 \pi^2 $, the kinetic mixing can arise from the multi-loop contribution, non-Abelian mixing, dynamical evolution, or the breaking of the dark charge conjugation symmetry $A' \rightarrow -A'$~\cite{Gherghetta:2019coi, Koren:2019iuv,Gan:2023wnp}.

\subsection{``Pure'' mCP}\label{subsec:pure_mcp}

Here, we consider a simple extension to SM, given the following Lagrangian:
\bea
\mathcal{L} \supset i \chibar (\slashed{\partial} - i g' q_\chi \slashed{B} + m_\chi) \chi - \frac{1}{4} B_{\mu \nu} B^{\mu \nu}.
\eea
While being agnostic of the UV theory, we assume here the new particle $\chi$'s small charge under $U(1)_Y$ NOT to be generated through kinetic mixing and hold in the fundamental UV theory. 
We thus call $\chi$ the ``pure'' mCP.

Most of the recent mCP studies focus on mCP from kinetic mixing, because of the apparent links to dark sectors and compatibility with GUT models. The ``pure'' mCP, on the other hand, also has great theoretical interests, as it provides indirect tests to versions of GUTs and string compactifications (see~\cite{Pati:1973uk,Georgi:1974my, Preskill:1984gd, Wen:1985qj, Shiu:2013wxa,Feng:2014eja} for detailed discussions.)

In terms of phenomenology, one may naively consider ``pure" mCP as effectively the case in which one takes $g_d\rightarrow 0$ for the kinetic-mixing mCP. However, the choice of taking $g_d\rightarrow 0$ or having a very small $g_d$ is not always self-consistent.
The matrix of the kinetic terms, which includes the kinetic-mixing term and the diagonal terms, has to be positive definite~\cite{Weinberg:1996kr, Burgess:2008ri}, which limits $\epsilon<1$. 
Under this condition, 
\bea
\epsilon<1 \quad \Rightarrow \quad g_d > e q_\chi 
\eea
which has important cosmological consequences (e.g., a large $q_\chi$ would also mean a large $g_d$ which makes $\chi \chibar \leftrightarrow A' A'$ is an efficient channel for energy transfer). Furthermore, it gives us a window to identify a region in which ``pure'' mCP is preferred while the mCP from kinetic mixing is disfavored, as discussed in the following sections. 

\vspace{-1mm}

\section{C\MakeLowercase{m}B cosmology}
\label{sec:cosmology}

\subsection{The reheating scenario}

Reheating is a process that, at the end of inflation, the inflaton $\phi$ decays and transfers the energy to the SM sector, populating and forming the thermal bath of SM particles~\cite{Kofman:1994rk, Kofman:1997yn}. When the decay rate of inflation $\Gamma_\phi \sim H$, most of the inflaton's energy is transferred to the SM radiation, therefore $\rho_\phi \sim  \rho_\rad$. The reheating temperature $T_\rh$ is defined as the temperature of the SM radiation when $\rho_\phi \sim  \rho_\rad$. There are many different reheating scenarios with different cosmological evolutions~\cite{Kofman:1997yn, Felder:1998vq,Felder:1998vq,Chung:1998rq, Giudice:2000ex,Tanin:2017bzm, Co:2020xaf,Fan:2021otj,Barman:2022tzk}. Generically, there is another temperature relevant to our discussion, which is the maximum temperature SM particles reach during the reheating process, denoted as $T_{\rm max}$.

In this work, we consider the scenario that $T_{\max} \simeq T_\rh$~\cite{Co:2020xaf, Barman:2022tzk, Frangipane:2021rtf,Bringmann:2021sth,Bhattiprolu:2022sdd}. 
This choice would generically lead to the minimal, and thus irreducible, amount of mCP produced during the reheating process. 
Obviously, one can also consider alternative scenarios in which $T_{\max}$ is larger than $T_\rh$, and they will generally lead to larger mCP production during the reheat process~\cite{Chung:1998rq, Giudice:2000ex, Drees:2017iod, Harigaya:2014waa, Harigaya:2019tzu, Garcia:2020eof,Garcia:2020wiy,Garcia:2021iag, Bernal:2022wck,Hooper:2023brf, Silva-Malpartida:2023yks,Becker:2023tvd}.
We emphasize that our consideration can also be extended to these scenarios. In fact, given a specific reheating scenario, one can determine the region of interest in the parameter space for an mCP test based on our method.

\subsection{``Pure'' CmB}
\label{subsec:pure_CmB}

As discussed in \Subsec{pure_mcp}, the ``pure'' mCP is directly charged under $U(1)_Y$ without coupling to a dark photon. The mCP can be produced through channels including $f \fbar \leftrightarrow \chi \chibar$ and $Z \leftrightarrow \chi \chibar$, where $f$'s and $Z$ are the SM fermions and Z boson from the reheating. 
Even though we do not require $\chi$ to be DM, the $\chi$ abundance $\Omega_\chi h^2$ in the late universe may exceed the currently observed DM abundance $\Omega_\DM h^2=0.12$ from these production channels. 
In addition, in the region where $\Omega_\chi \gtrsim 0.4\%\;\Omega_\DM$, CMB angular power spectrum can be distorted by the mCP-baryon scattering when $q_\chi$ is sizable~\cite{WMAP:2012fli, Aghanim:2018eyx, ACT:2020gnv}. These are considered overabundance and mCP-baryon interaction constraints for the ``pure'' CmB.

There are two processes, $f \fbar \leftrightarrow \chi \chibar$ and $Z \leftrightarrow \chi \chibar$, that are relevant for the ``freeze-in" and ``freeze-out" of the mCPs.
Freeze-in simply means that mCP is produced from the SM bath~\cite{Hall:2009bx}. Freeze-out means mCP annihilation to the SM particle would become inefficient in the late universe, and the $\Omega_\chi$ is fixed after the ``freeze-out" of these processes.
We consider the Boltzmann equation of $\chi$ and the SM bath to calculate the abundance of $\chi$, detailed in \Appx{pure_mcp_abundance}. 
Even though our results are derived based on solving the Boltzmann equation, we also explain them with analytical estimations below. The reheating temperature, i.e., $T_\rh$, is important for the abundance of $\chi$ produced, thus we discuss the analytical estimation of how the $\chi$ abundance depends on $T_\rh$. 

First, we consider $m_\chi \lesssim T_\rh$. In this case, the region of the overproduction is enveloped by the freeze-in and freeze-out curves. 
To estimate the mCP abundance in the freeze-in region, one can utilize the equation
\bea
\label{eq:YX_FI}
Y_{\chi}^\fii \sim q_\chi^2 \alphaem^2 \frac{\mpl}{T}, \;\: T \gtrsim m_\chi.
\eea

In this case, the dominant part of mCP is produced at $T\sim m_\chi$ and we have $Y_{\chi} \sim q_\chi^2 \alphaem^2 \mpl/m_\chi$, thus
\bea
\Omega_\chi h^2 \sim 0.12 \left( \frac{q_\chi}{10^{-11}} \right)^2.
\eea
This shows that along the relic abundance contour, $q_\chi$ is roughly independent of $m_\chi$.
The production is determined by the Boltzmann equation in \Appx{pure_mcp_abundance}, and the numerical $\Omega_\chi h^2$ evolution is shown in \Fig{mcp_contour_high_Trh}. All the relevant channels and the evolution of the effective degree of freedom for entropy density, $g_{*,S}$ are taken into account properly. Therefore, the mCP production process is classified as the ``IR freeze-in". 

To estimate the mCP abundance in the freeze-out region, we utilize
\bea
\label{eq:mcp_yield_fo}
Y_{\chi}^\fo \sim \frac{1}{q_\chi^2 \alphaem^2} \frac{m_\chi}{\mpl},
\eea
where $Y_{\chi,\fo}$ is the yield of the mCP after freeze-out. 
Substituting \Eq{mcp_yield_fo} into $\Omega_\chi h^2 \sim Y_{\chi}^\fo \, m_\chi /T_\eq$, we have 
\bea
\Omega_\chi h^2 \sim 0.12 \left( \frac{10^{-2}}{q_\chi} \right)^2 \left( \frac{m_\chi}{1\GeV} \right)^2.
\eea
This indicates that for the freeze-out curve of mCP, $q_\chi \propto m_\chi$ approximately, as we will show later in the plots.

For $m_\chi > T_\rh$. For the mCP in the freeze-out region, if the freeze-out temperature $T_\fo < T_\rh$, mCP cannot reach the thermal equilibrium with the SM bath. Therefore the freeze-out curve is cut off at $m_\chi \gtrsim x_\fo T_\rh$ where $x_\fo\equiv m_\chi/T_{\rm FO} \sim 20$. 
For the freeze-in curve, the rate of mCP's interaction with the SM bath is suppressed by $(m_\chi/T_\rh) \exp(-2m_\chi/T_\rh)$. So the freeze-in curve holds the approximate relation:
\bea
q_\chi  \approxprop   \left( \frac{T_\rh}{m_\chi} \right)^{1/2} \exp\left(\frac{m_\chi}{T_\rh}\right),
\eea 
which explains the exponential tilting up of the curve in the region $m_\chi > T_\rh$. 
Because of the exponential suppression, the dominant part of the mCPs is produced near $T\sim T_\rh$ in this case, which is known as the UV freeze-in~\cite{Elahi:2014fsa,UV_freezein}.

In the left panel of \Fig{bounds_Trh}, we plot the overproduction constraints of mCP depending on the reheating temperatures in different colors from $5\,\MeV$ to $100\,\GeV$ and above. 
From the plot, one can find that smaller $T_\rh$ leads to weaker overproduction constraint.
Because of the precision measurement of the light element abundance, the reheating should happen before the BBN, which sets $T_{\rh} \gtrsim 5\,\MeV$~\cite{Kawasaki:1999na, Kawasaki:2000en, Hannestad:2004px, Ichikawa:2006vm, deSalas:2015glj, Hasegawa:2019jsa}. We thus use the purple-shaded region to denote the overproduction constraint when $T_\rh = 5\, \MeV$, which cannot be alleviated by further lowering $T_\rh$.
We provide deviation of \Eq{YX_FI} and (\ref{eq:mcp_yield_fo}) in the \Appx{pure_mcp_abundance}. The dips of the freeze-in and freeze-out curves around $\m_\chi \sim m_Z/2$ are because of the $Z$-resonances, which are also discussed in \Appx{pure_mcp_abundance}.

\subsection{CmB from kinetic mixing}
\label{subsec:CmB_Kinetic_Mixing}
\vspace{-3mm}

For the mCP from kinetic mixing, the mCPs produced from the SM bath can annihilate to the massless dark photons through $\chi \chibar \rightarrow A' A'$. 
As the dark radiation consisting of massless $A'$ is populated, the expansion rate of the Universe can be affected, changing the effective number of relativistic degrees of freedom $N_\eff$. Such a change is denoted as $\Delta N_\eff$, which can be constrained by BBN and CMB measurements~\cite{Vogel:2013raa, Adshead:2022ovo, Luo:2020sho, Luo:2020fdt}.
The reheating temperature can significantly affect the population of $A'$, changing the $\Delta N_\eff$ as a consequence. 
In this section, we give the condition where $\alpha_d$ is large enough, and the majority of the $\chi$ entropy is dumped to $A'$ as $\chi$ becomes non-relativistic. Then, we estimate the change of $\Delta N_\eff$ constraints based on different reheating temperatures.

First, we consider the situation that $m_\chi\lesssim T_\rh$.
We use $n_\chi^\fii$ to denote the mCP's number density from the freeze-in; one has
\bea\label{eq:thermalization_cond}
n_\chi^\fii \sim q_\chi^2 \alphaem^2 \mpl T^2 
\eea
from \Eq{YX_FI}. 
We consider $\langle \sigma v \rangle_\dth$, the dominant thermal cross-section of the channels $\chi \chibar \leftrightarrow A' A'$, $\chi A' \leftrightarrow \chi A'$, which thermalize the dark sector and populate $A'$, 
$
\langle \sigma v \rangle_{\dth} \sim \alpha_d^2/T^2,
$
and $\chi A' \leftrightarrow \chi A' A'$ can keep the chemical potential of $A'$ to be zero.
The condition for $\chi$ to effectively transfer its entropy to dark radiation $A'$ is
\bea\label{eq:condition}
\frac{n_\chi^\fii \langle \sigma v \rangle_{\dth}}{H} \sim q_\chi^2 \alphaem^2 \alpha_d^2 \left(\frac{\mpl}{T}\right)^2 \gg 1.
\eea
For example, for $q_\chi\sim 10^{-11}$ and $T \sim 100\,\GeV$, with $\alpha_d \gg 10^{-4}$, condition (\ref{eq:condition}) can be satisfied. 

For this section, we consider $\alpha_d \sim \alphaem$, as a natural choice motivated by the mirror symmetry between the SM sector and the dark sector~\cite{Chacko:2005pe,Chang:2006ra,Craig:2015pha,Dunsky:2019api, Batell:2019ptb,Liu:2019ixm, Batell:2022pzc}.
Once $A'$ is populated, $\chi$ will start exiting the SM bath as $\chi$ and $A'$ form a dark sector with temperature $T_d$, as \bea
\frac{ \langle \sigma v \rangle_{\chi \chibar \leftrightarrow A' A'} }{ \langle \sigma v \rangle_{\chi \chibar \leftrightarrow f \fbar} } \sim \frac{\alpha_d^2}{ q^2_\chi \alphaem^2} \gg 1,
\eea 
The aforementioned $
\chi-A'$ channels will keep the dark sector thermalized until $\chi$ becomes fully non-relativistic, and all the $\chi$ entropy will be transferred to the $A'$ sector.
As a result, we have
\bea\label{eq:Neff_energy_inj}
\Delta N_\eff \sim q_\chi^2 \alphaem^2 \frac{\mpl}{m_\chi}
\eea
in the region where $\Delta N_\eff \lesssim 1$ and $g_*$ has small variation during the energy injection. 
Based on \Eq{Neff_energy_inj}, the $\Delta N_\eff$ constraint boundary is 
\bea
\label{eq:Neff_const}
q_\chi \sim 10^{-7} \left(\frac{m_\chi}{1\,\GeV}\right)^{1/2} \left( \frac{\Delta N_\eff}{0.3} \right)^{1/2}.
\eea
One can find the detailed derivation of \Eq{Neff_energy_inj} in \Appx{energy_inj}.

For $ m_\chi> T_\rh$, the energy injection rate of SM bath to the dark sector is suppressed exponentially. In this case, the $\Delta N_\eff$ contour behaves as
\bea
q_\chi \approxprop  \exp\left( \frac{m_\chi}{T_\rh} \right).
\eea

In the right panel of \Fig{bounds_Trh}, we show the $\Delta N_\eff$ constraints on $A'$ (from mCP $\chi$) depending on the reheating temperatures. Plotted in different colors, these curves denote the cases where $T_\rh$ varies from $5\,\MeV$ to 1\,\GeV\; and above. 
Again, given the current understanding that $T_\rh$ should be larger than the temperature of BBN, we shade the constraint corresponding to $T_\rh$ = 5 MeV with a purple color to indicate that this region of constraint cannot be easily alleviated by lowering the reheating temperature.

As one can see in the right panel of \Fig{bounds_Trh}, as $T_\rh \gtrsim 1\,\GeV$, the current $\Delta N_{\eff} \le (0.3)_{\rm Planck}$ constraint from Plank measurement~\cite{Aghanim:2018eyx} stop moving to the right, meaning that, for $T_\rh$ = 1 to 100 GeV and above, the corresponding curve does not change.
This is because of two reasons.
First, for $m_\chi \gtrsim$ GeV, there is a maximum $\Delta N_\eff$ value from the dark radiation given by the degree of freedom of $A'$, $g_{A'}$, since $\chi$ will become non-relativistic during BBN.
If the SM bath and the dark bath lose thermal contact before QCD phase transition, the $\Delta N_\eff$ contribution of dark radiation $A'$  is diluted by the change of SM degrees of freedom $g_*$ as
\bea
\label{eq:DeltaNeff_QCD}
\Delta N_\eff \lesssim g_{A'}\:\frac{4}{7} \left(\frac{g_{*,S}(T \ll T_\QCD)}{g_{*,S}(T\gg T_\QCD)} \right)^{4/3} \simeq 0.1,
\eea
where $g_{A'}=2$. \Eq{DeltaNeff_QCD} can be derived from the entropy conservation, $T\simeq T_d$ before the QCD phase transition, and mCP becomes non-relativistic during this period.
When the mCP becomes non-relativistic, it can heat up the SM bath and $A'$ dark radiation unevenly to raise $T_d/T$ slightly above 1, but it does not change the conclusion qualitatively~\cite{Vogel:2013raa, Adshead:2022ovo}.

Furthermore, if one considers $\Delta N_{\eff} \le (0.06)_\text{CMB-S4} $ corresponding to future CMB-S4 projection~\cite{Abazajian:2019eic,CMB-S4:2022ght}, the constraints can continue to extend to the right with higher $T_\rh$.

In this section, we discuss the case that $\alpha_d$ is large enough to keep $\chi$ and $A'$ fully thermalized. Generically, one can consider the case of a small $\alpha_d$ that $\chi$ and $A'$ are only partially thermalized or not thermalized, which would be an interpolation of the considerations in \Subsec{pure_CmB} and \Subsec{CmB_Kinetic_Mixing}. Our new $T_\rh$-dependent results of the $\Delta N_\eff$ constraints provide an opportunity to test the reheating scenarios with dedicated mCP searches, discussed in the following section.

\section{Results and discussions}
\label{sec:results}

\subsection{Testing reheating scenarios}

Here, we discuss the implication of the cosmological ``constraints''. We point out that they are actually target regions to test different reheating scenarios. More specifically, if one finds an mCP within these constraint regions, one can set an upper limit on $T_{\rm rh}$ in these scenarios.
As prime examples, given the beam-production accelerator searches are independent of the mCP abundance, they become tools to test the reheating scenarios. The dedicated mCP searches are especially powerful since it is harder to replicate the mCP signatures in these experiments with other SM or BSM particles.

\begin{figure*}[t]
    \centering
    \includegraphics[width=0.45
\textwidth]{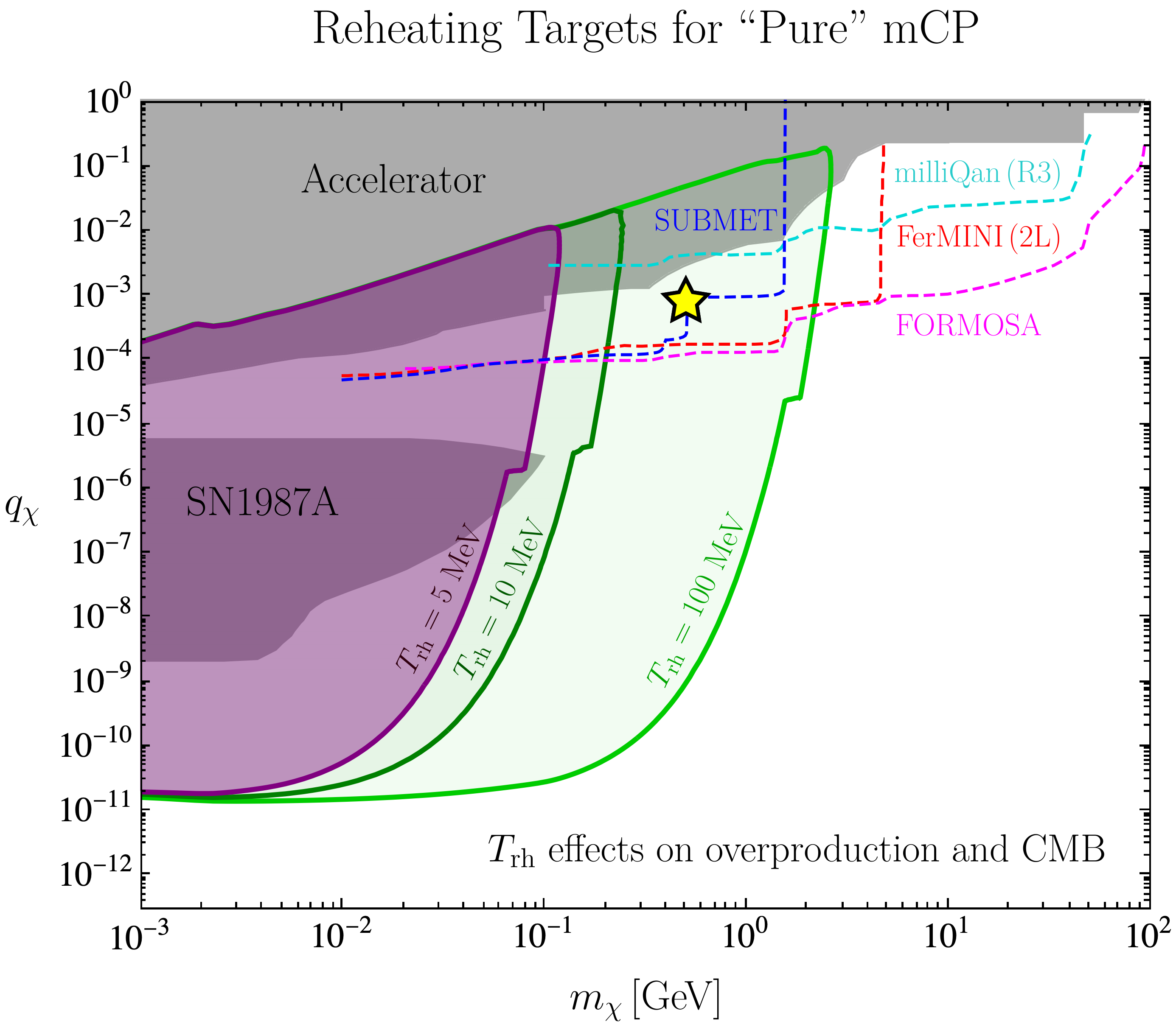} \qquad
\includegraphics[width=0.45\textwidth]{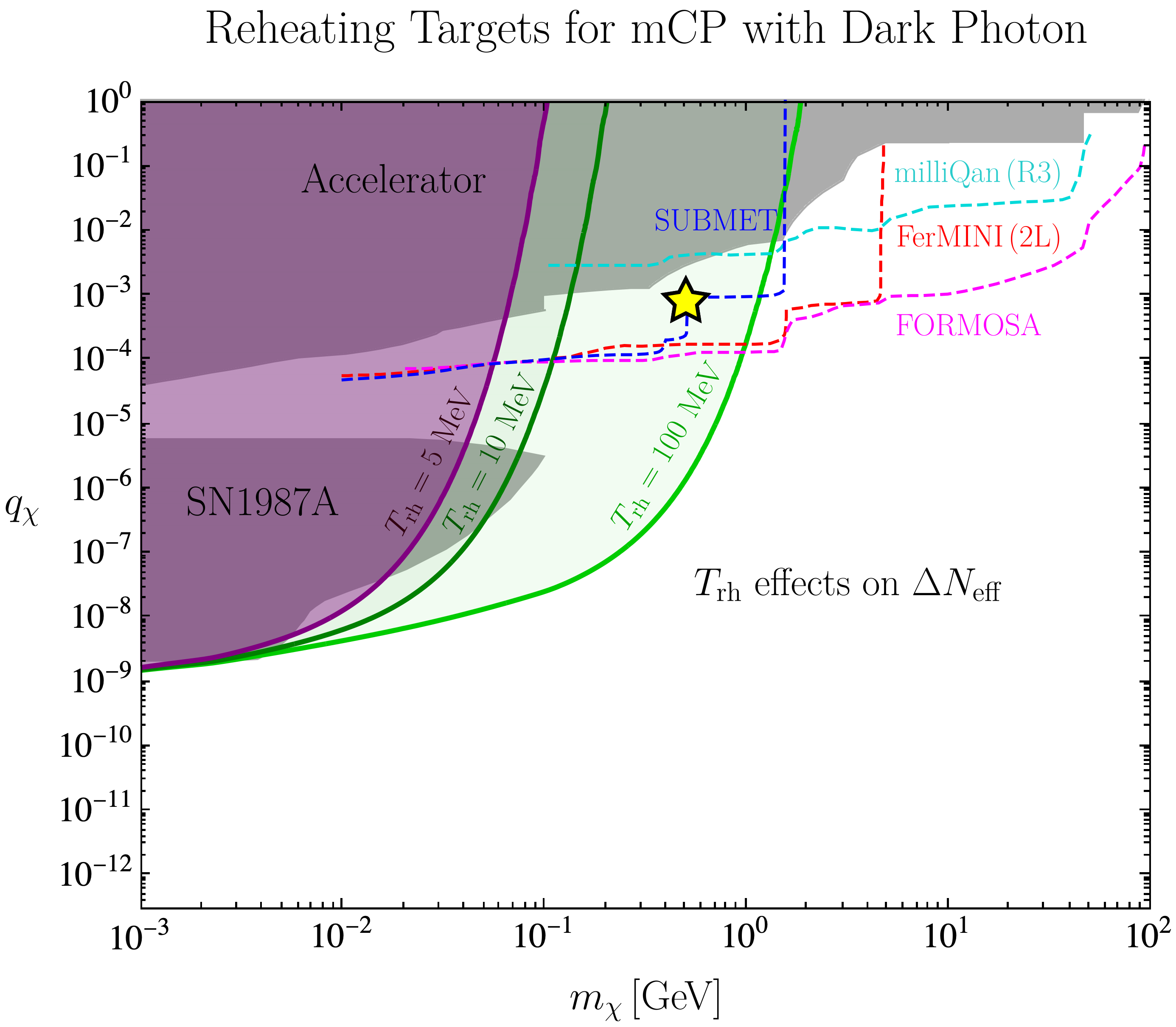}
\caption{
{\bf Left}: The reheating temperature target regions for ``pure'' mCPs. The regions enveloped by the light and dark green solid lines are the combinations of constraints on mCP overproduction and mCP-baryon interactions, given that $T_\rh = 10\,\MeV, 100\,\MeV$. 
The purple-shaded region combines the irreducible overproduction and mCP-baryon interaction constraints when $T_\rh$ saturates the lower bound. {\bf Right}: The reheating temperature test is based on the mCP from kinetic mixing. The regions enveloped by the dark and light green lines are the dark photon's $\Delta N_\eff$ constraints from $\chi \chibar \rightarrow A' A'$, given that $T_\rh = 10\,\MeV, 100\,\MeV$. The purple-shaded region is the irreducible $\Delta N_\eff$ constraint when $T_\rh$ saturates the lower bound. 
In both panels, we use yellow stars to label the parameter point $m_\chi=0.5 \,\GeV$, $q_\chi=8 \times 10^{-4}$ as a demonstration that, if one finds an mCP at this parameter point, one can set an upper bound on $T_\rh<100\:\MeV,$ for both ``pure" and kinetic mixing mCPs.
}
\label{fig:constraints2}
\end{figure*}

In \Fig{constraints2}, we further demonstrate this point by choosing a parameter point testable by near-future mCP-dedicated searches.

\begin{figure*}[t]
    \centering
    \includegraphics[width=0.45
\textwidth]{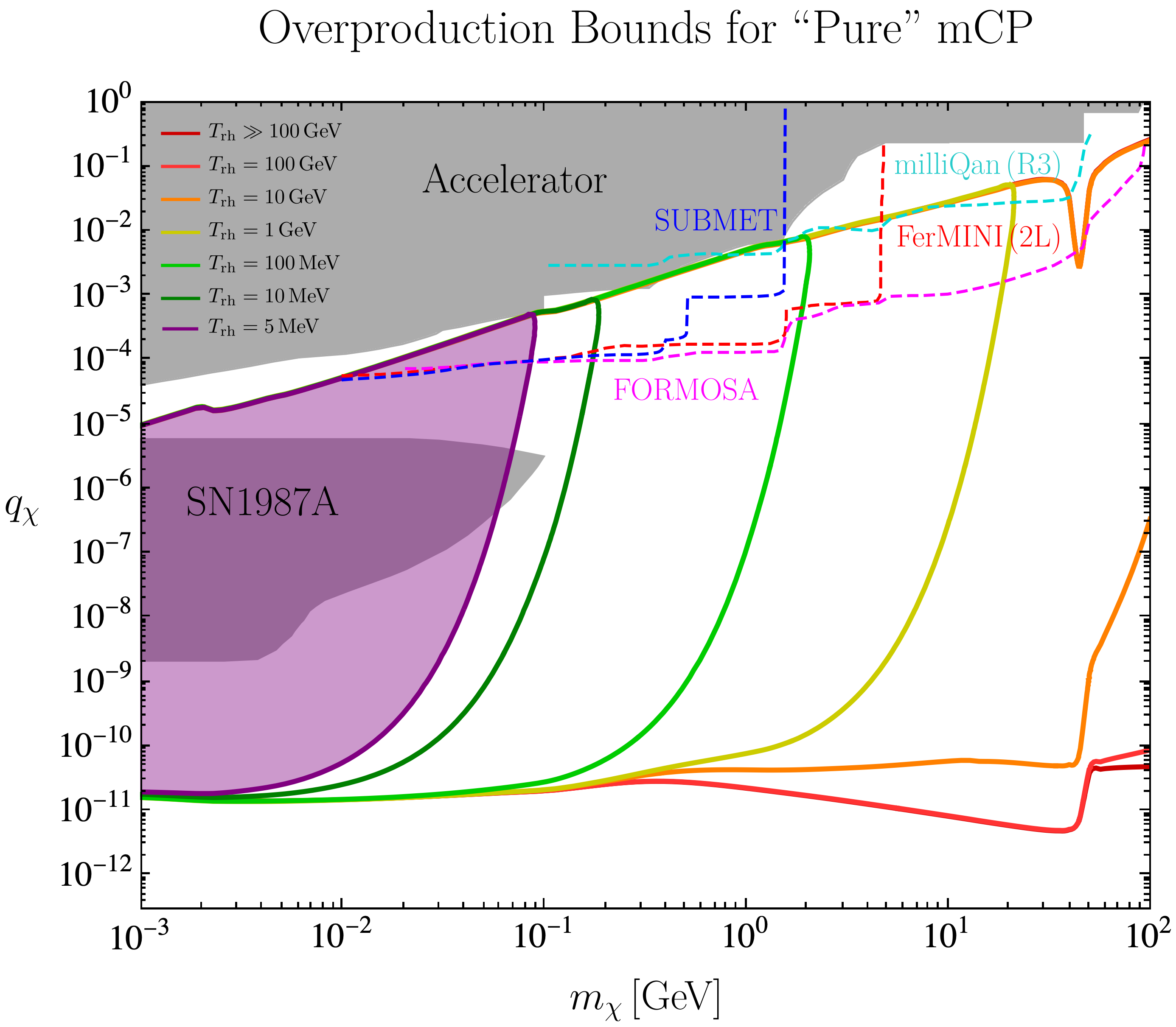} \qquad
\includegraphics[width=0.45\textwidth]{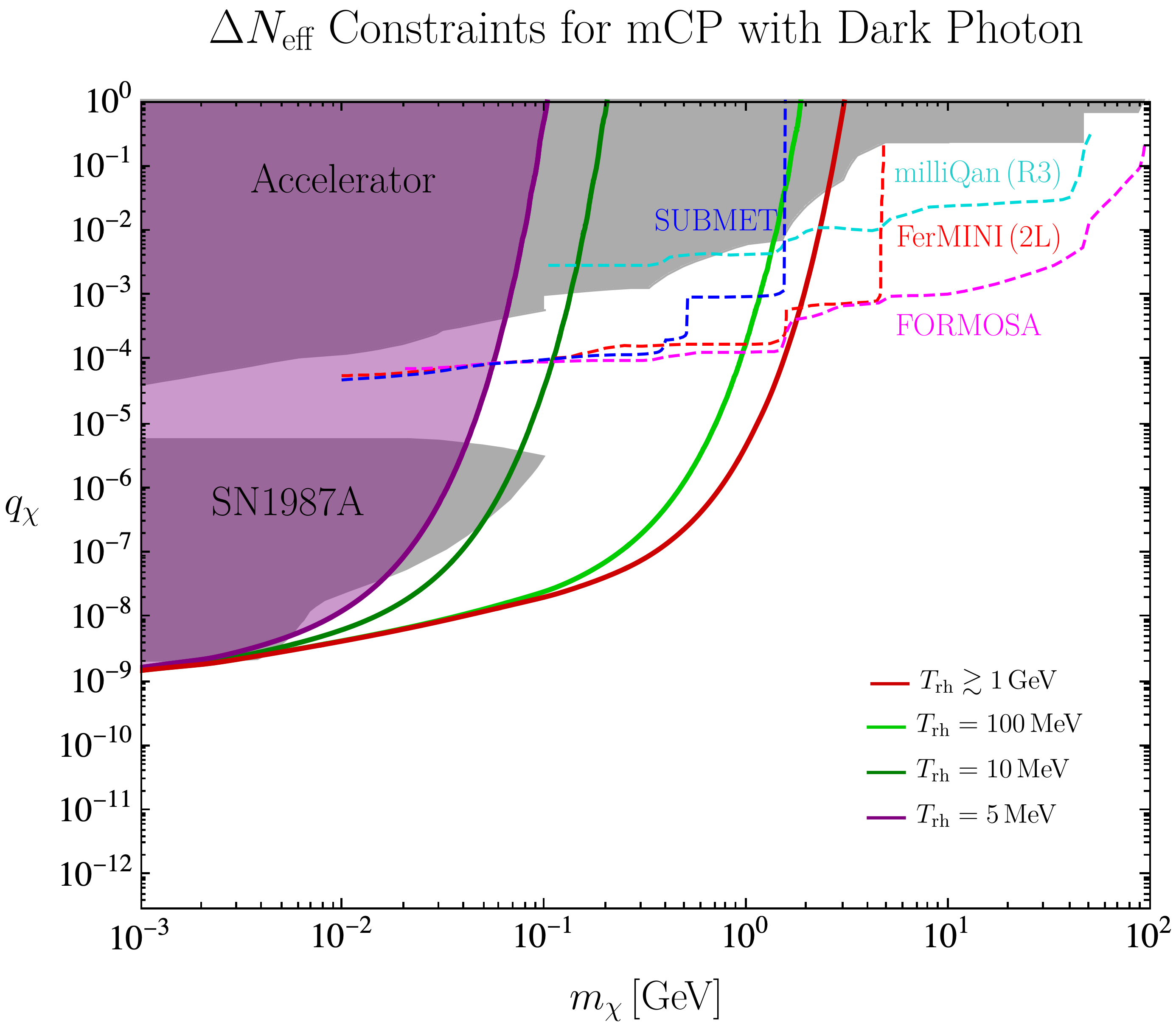}
\caption{{\bf Left}: The overproduction constraints of ``pure'' mCP with different reheating temperatures ranging from 10 MeV to above 100 GeV. The purple-shaded region is the irreducible constraint based upon the mCP's overproduction when $T_\rh$ saturates the lower bound set by BBN. {\bf Right}: The $\Delta N_{\rm eff}$ constraints for the dark photon of the kinetic-mixing mCP with different reheating temperatures ranging from 10 MeV to above GeV. The purple-shaded region is again the irreducible $\Delta N_\eff$ constraint when $T_\rh$ saturates the lower bound.  
In both panels, 
we also plot the dedicated mCP searches relevant to these regions, including milliQan~\cite{Ball:2016zrp, milliQan:2021lne}, FerMINI~\cite{Kelly:2018brz}, SUBMET~\cite{Choi:2020mbk, Kim:2021eix}, and FORMOSA~\cite{Foroughi-Abari:2020qar}. 
}
\label{fig:bounds_Trh}
\end{figure*}
\clearpage

First, we consider the ``pure'' mCP case. In the left panel of \Fig{constraints2}, the shaded region enveloped by the darker and lighter green lines denotes the combinations of the constraints from ``pure'' mCPs' overproduction and the CMB constraints on the mCP-baryon scattering with $T_\rh = 10 \,\MeV, 100\,\MeV$, respectively. Note that, for the CMB constraint on the mCP-baryon scattering, we plot the $\Omega_\chi h^2 = 0.4\% \times \Omega_\DM h^2$ contour~\cite{Dubovsky:2003yn, Dolgov:2013una, dePutter:2018xte, Kovetz:2018zan,Xu:2018efh,Buen-Abad:2021mvc}. 
If the dedicated accelerator experiments such as milliQan, FerMINI, SUBMET, and FORMOSA find the signal of mCP between $T_\rh=10 \, \MeV$ and $T_\rh = 100\, \MeV$ contours~(yellow star), we can set the upper bound on the reheating temperature to be $T_\rh \lesssim 100 \, \MeV$.

For the kinetic-mixing mCP, a similar region is shown in the right panel of \Fig{constraints2}. The region of interest enclosed by darker and lighter green curves denote the boundaries of the $\Delta N_\eff$ constraints where $T_\rh = 10\,\MeV, 100\,\MeV$, respectively. If the dedicated searches find an mCP between these two lines~(demonstrated by a yellow star), we can set the upper bound on the reheating temperature $T_\rh \lesssim 100\,\MeV$. Note that the yellow stars in both left and right panels are chosen at the same parameter point $m_\chi=0.5 \,\GeV$, $q_\chi=8 \times 10^{-4}$ as an illustration, meaning one can test the reheating scenarios for both types of mCPs. 
One can probe other reheating temperatures for the ``pure'' mCP and kinetic mixing mCP using the regions plotted in \Fig{bounds_Trh}

In \Fig{four_regions}, we show the $\Delta N_\eff \le (0.3)_{\rm Planck}$ constraint assuming $T_\rh \gtrsim  1\, \GeV$ with a solid red curve (for $\Delta N_\eff \le (0.3)_{\rm Planck}$, as long as $T_\rh \gtrsim  1 \,\GeV$, the constraint curve does not change by considering a higher reheating temperature), as well as the CMB-S4 projection for the $\Delta N_\eff \le (0.06)_\text{CMB-S4}$ constraint corresponding to $T_\rh \gg$ 100 GeV with a dot-dashed red curve. We derive these curves, and the results are consistent with~\cite{Adshead:2022ovo}.  
For the kinetic mixing mCP, we identify a region in which dedicated accelerator searches, combined with the future CMB-S4 measurement~\cite{Abazajian:2019eic,CMB-S4:2022ght} can probe $T_\rh$ larger than 1 GeV, labeled by the green star and enclosed by the solid and dot-dashed red curves. We only show \Fig{four_regions} above $m_\chi = 10\,\MeV$, because, below this mass, there is an additional constraint from BBN and CMB based on the $\chi$ energy density and energy transfer with the SM bath~\cite{Boehm:2012gr, Krnjaic:2019dzc, Giovanetti:2021izc}.

\subsection{Theoretically motivated regions: \\ 
how to differentiate the two types of mCPs?}
\label{subsec:theory}

The consideration of CmB cosmology gives us windows to probe theoretically motivated regions and the origins of mCPs.
Here, we start with identifying a region where we can potentially search for the ``pure'' mCP. The ``pure'' mCP is not compatible with versions of GUTs and string compactifications and can be viewed as an indirect test for these theories~\cite{Pati:1973uk,Georgi:1974my, Preskill:1984gd, Wen:1985qj, Shiu:2013wxa,Feng:2014eja}.
One may tune the mCP-dark photon coupling $g_d$ close to zero as an attempt to allow the kinetic mixing mCP to behave like ``pure" mCP cosmologically.
However, there is a lower bound of $g_d$ for a given $q_\chi$ due to the requirement that the matrix of the kinetic terms should be positive definite, and thus $\epsilon<1$, discussed in \Subsec{pure_mcp}.

\begin{figure}[t]
\includegraphics[width=8.5cm]{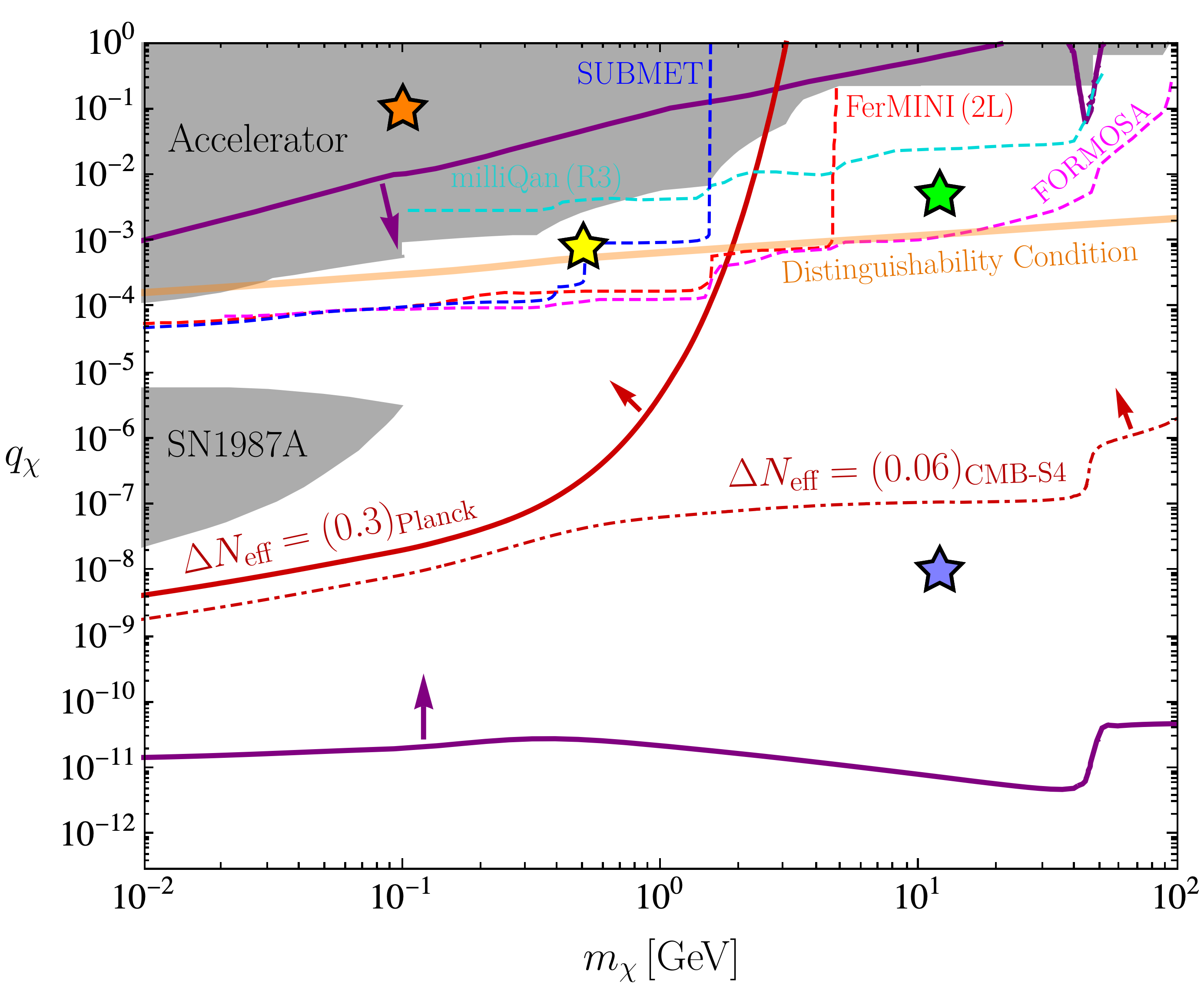}
\caption{Targets for the mCP searches from cosmological and theoretical considerations.
In this plot, we consider $T_\rh\gg 
100\;\GeV$, and the directions of the arrows point to the regions of constraints.
The {\bf yellow star} labels the region enclosed by the two ``pure" mCP constraint curves (purple) and the kinetic-mixing $\Delta N_\eff$ constraint curve (solid red). One can test the reheating scenario and temperature for both mCP scenarios in this region. Note that this yellow star is chosen at the same parameter point, $m_\chi=0.5 \,\GeV$, $q_\chi=8 \times 10^{-4}$, as the yellow stars in \Fig{constraints2}.
The {\bf orange star} labels the region which favors the ``pure" mCP particle, which is above both the upper purple curve and the orange curve, which is the ``Distinguishability Condition" discussed in \Subsec{theory}. This region is already explored by previous accelerator experiments.
The {\bf green star} is the region enclosed by future CMB-S4 sensitivity to $\Delta N_\eff\le 0.06$, which will allow us to study reheating temperatures above 1 GeV for the kinetic-mixing mCP, as discussed in detail in \Subsec{CmB_Kinetic_Mixing}.
One can see that the combination of DM overproduction, mCP-baryon interaction, and future $\Delta N_\eff$ constraints can allow us to probe the reheating temperatures up to 100 GeV and above for both mCP scenarios.
The {\bf green star} and {\bf blue star} label the regions where kinetic-mixing mCP are currently favored.
Since the future CMB-S4 constraint will not cover the blue star, it will remain to favor kinetic-mixing mCP.
}
\label{fig:four_regions}
\end{figure}

More specifically, for kinetic-mixing CmB to behave like ``pure" CmB, one can tune the $g_d$ small enough that the $\chi \chibar \rightarrow A' A'$ channel is inefficient during the cosmic evolution, i.e.,
$n_\chi^\eq \langle \sigma v \rangle_{\chi \chibar \rightarrow A' A'}/H \lesssim 1$, which would give an upper bound on $g_d$. Specifically, this non-thermalization condition of the mCP with a dark photon is $g_d\lesssim ( 16 \pi^2 m_\chi/\mathcal{F}\mpl )^{1/4}$, where $\mathcal{F} \simeq 375/(16\pi^3 e^{5/2} g_*^{1/2} )$. 
On the other hand, $\epsilon<1$ requires $g_d > e q_\chi$. Considering these two inequalities for $g_d$, we can roughly determine that 
\bea
q_\chi \gtrsim \frac{1}{\alphaem^{1/2}} \left(\frac{m_\chi}{\mathcal{F}\mpl}\right)^{1/4} 
\label{eq:epsilon<1}
\eea
is the region in which the two inequalities CANNOT be satisfied simultaneously.
We label \Eq{epsilon<1} with the orange curve in \Fig{four_regions}. 

Combining these discussions, we can determine three conditions for the ``pure" mCP favored region:
\begin{itemize}
    \item [1.] $q_\chi \gtrsim \frac{1}{ \alphaem^{1/2} } \left( \frac{m_\chi}{\mathcal{F}\mpl} \right)^{1/4}.$
    \item [2.] Constrained by the $\Delta N_\eff$ bound on kinetic-mixing mCP.
    \item [3.] Allowed by the overproduction and mCP-baryon scattering bounds on ``pure'' mCPs.
\end{itemize}
In \Fig{four_regions}, we labeled this region, which is the area above the upper purple curve, with an orange star. One can see that existing accelerator searches have successfully excluded this region. 
In Fig.~\ref{fig:bounds_Trh}, \ref{fig:constraints2}, \ref{fig:four_regions}, the accelerator constraints are compilations of \cite{Davidson:2000hf,Prinz:1998ua,Badertscher:2006fm,CMS:2012xi,Jaeckel:2012yz,Magill:2018tbb,ArgoNeuT:2019ckq,Marocco:2020dqu,milliQan:2021lne,SENSEI:2023gie}.

Note that \Eq{epsilon<1} is based on the assumption that $T_\rh\gg 100\,\GeV$. The condition has to be modified with low reheating temperatures. 

Our cosmological consideration also highlights the regions where kinetic-mixing  mCP with sizable coupling to a dark photon is favored.
More specifically, if an mCP is constrained by the over-production and mCP-baryon scattering (for ``pure" CmB from \Subsec{pure_CmB}), but not by the $\Delta N_\eff$ bound (for kinetic-mixing CmB from \Subsec{CmB_Kinetic_Mixing}), the mCP is favored to have significant dark photon coupling. In \Fig{four_regions}, the regions favoring kinetic-mixing mCPs are labeled by green and blue stars. The major difference between the green-star and blue-star regions is simply that the green-star region can be covered by future CMB-S4 measurements, which can help test the dark-photon hypothesis.
It is hard to probe the blue-star region with future accelerator searches due to the limitation of luminosity. Searches like direct-detection experiments, although highly dependent on the mCP relic abundance, can explore this region in the future. 

\section{Conclusion and outlook}

In summary, we consider the irreducible cosmic millicharged background (CmB) for both ``pure" and kinetic-mixing mCPs, and identify regions of parameter spaces where future experiments can test the reheating scenarios and temperatures.
The consideration of using dark-sector searches as probes of reheating scenarios, discussed in this paper, can be extended to other BSM particles.
We also provide cosmological studies to differentiate the two major mCP scenarios, as both have distinct and profound theoretical motivations given their UV origins.

\section*{Acknowledgement}

We thank Gary Shiu, Joshua T. Ruderman, Christopher S. Hill, Matthew D. Citron, Kevork N. Abazajian, Cyril Creque-Sarbinowski, Keisuke Harigaya, Andrew J. Long, and Jessie Shelton for valuable discussions. We also thank Zhen Liu for commenting on the draft. Y.-D.T. is supported by the U.S. National Science Foundation (NSF) Theoretical Physics Program, Grant No.~PHY-1915005. Y.-D.T. also thanks the Institute for Nuclear Theory at the University of Washington for its kind hospitality and stimulating research environment. This research was supported in part by the INT's U.S. Department of Energy grant No.~DE-FG02-00ER41132. This work was partially performed at the Aspen Center for Physics, supported by National Science Foundation grant No.~PHY-2210452. This research was partly supported by the National Science Foundation under Grant No.~NSF PHY-1748958. This document was partially prepared using the resources of the Fermi National Accelerator Laboratory (Fermilab), a U.S. Department of
Energy, Office of Science, HEP User Facility. Fermilab is
managed by Fermi Research Alliance, LLC (FRA), acting under Contract No. DE-AC02-07CH11359.
This work was supported in part through the NYU IT High-Performance Computing resources, services, and staff expertise.

\appendix

\section{``Pure'' mCP abundance from the Boltzmann equation}
\label{appx:pure_mcp_abundance}

\begin{figure*}[t]
    \centering
    \includegraphics[width=0.45
\textwidth]{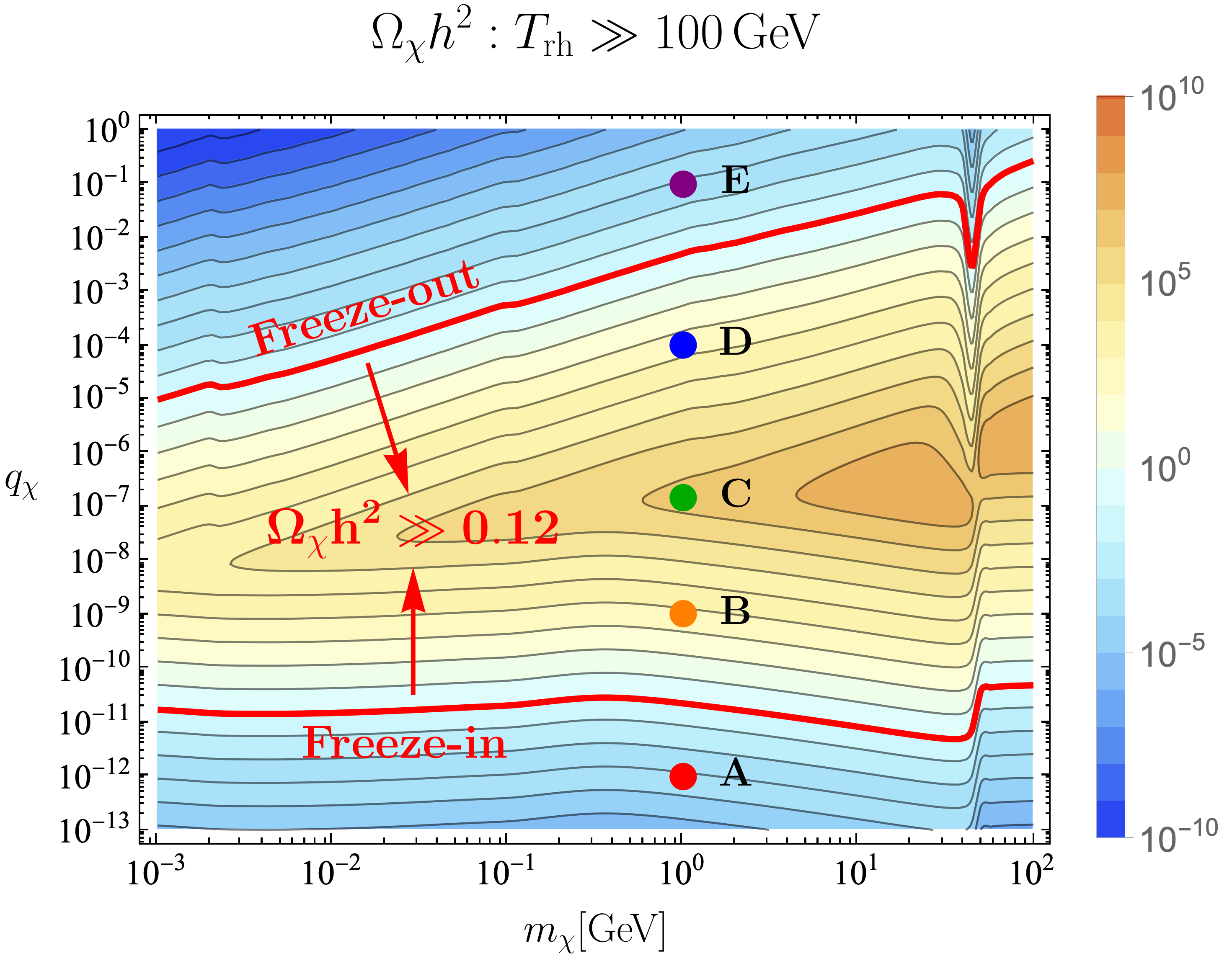} \qquad
\includegraphics[width=0.45\textwidth]{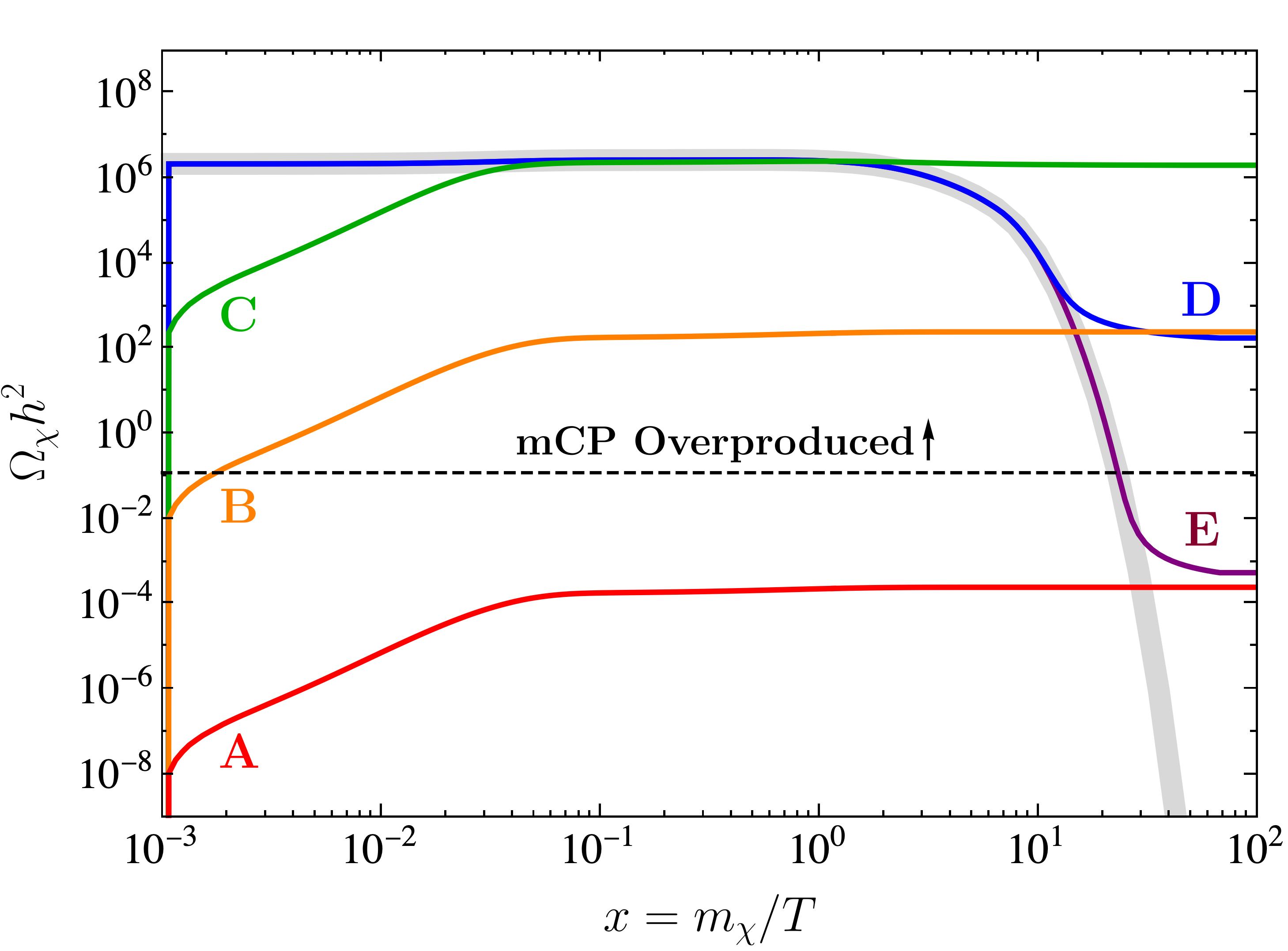}
\caption{ {\bf Left}: For $T_\rh \gg 100 \, \GeV$, the contours of the ``pure'' mCP's late-time ($x\equiv m_\chi/T > 100$) relic abundances. The lower\,(upper) red curve is the freeze-in\,(freeze-out) curve. $\Omega_{\chi} h^2 \gg 0.12$ in the region enveloped by these two curves. 
The points B, C, and D are related to the overproduction of the mCPs, while A and E are related to the case in which mCPs are the subcomponent of the dark matter.
{\bf Right}: The cosmological evolution of the abundances of ``pure'' mCPs. The abundances evolve from zero values, corresponding to the irreducible productions.
Each colored curve in the right panel corresponds to the relic abundance evolution of the colored point in the left panel, with $m_\chi=1\;\GeV$ and different $q_\chi$.
Curves A and B correspond to small ${q_\chi}'s$, with the abundances gradually increasing and keeping constant after $T \sim m_\chi$; this kind of evolution is known as freeze-in. Curves D and E correspond to large ${q_\chi}'s$, and mCPs reach the thermal equilibrium at $T \simeq T_\rh$, then go along the thermal distribution until they decouple from the chemical equilibrium; this is known as the freeze-out scenario. 
Curve C is in between the freeze-in and freeze-out scenarios. All the curves are from solving the Boltzmann equation \Eq{boltz_pure_mcp}.
}
\label{fig:mcp_contour_high_Trh}
\end{figure*}

\begin{figure*}[t]
    \centering
    \includegraphics[width=0.45
\textwidth]{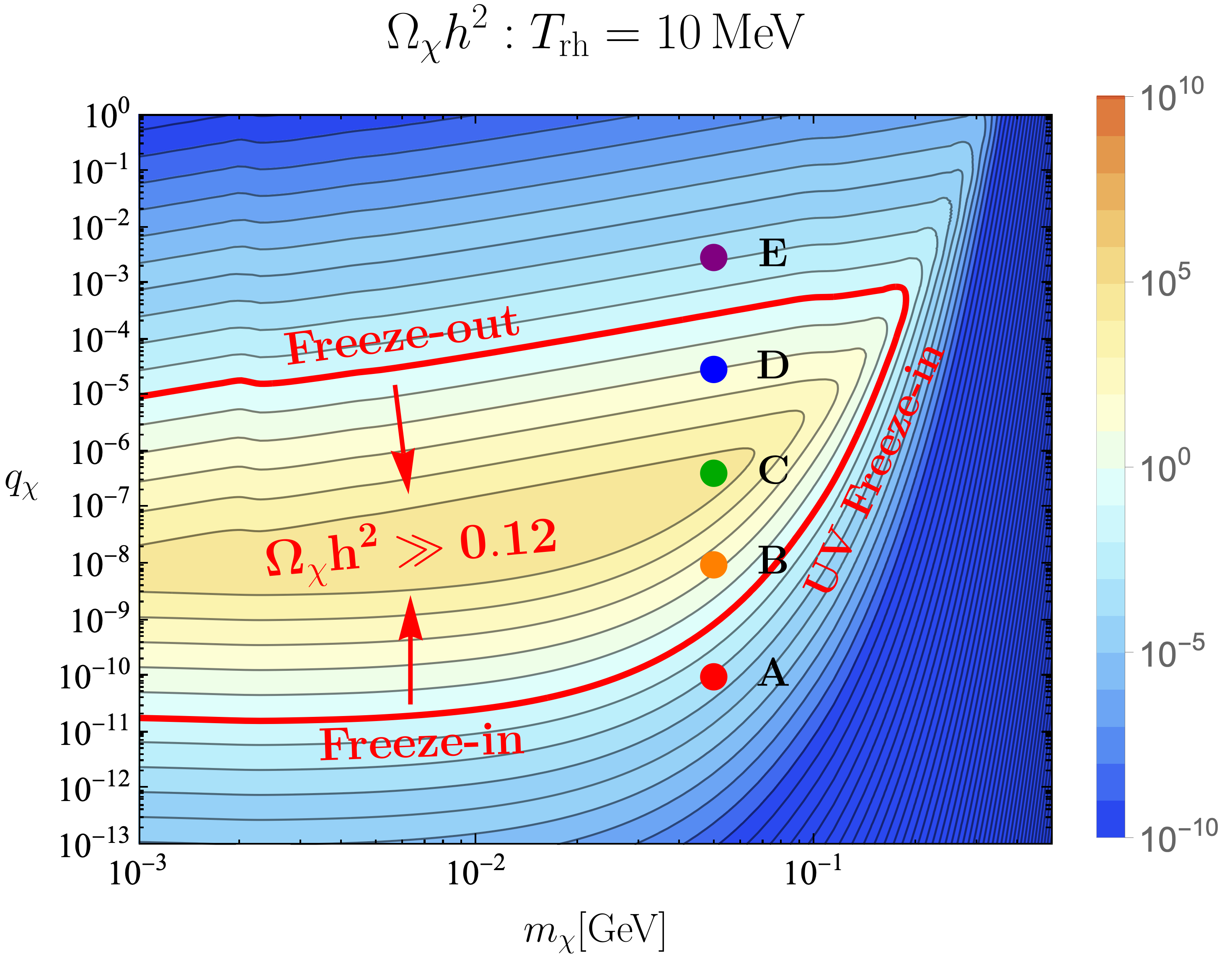} \qquad
\includegraphics[width=0.45\textwidth]{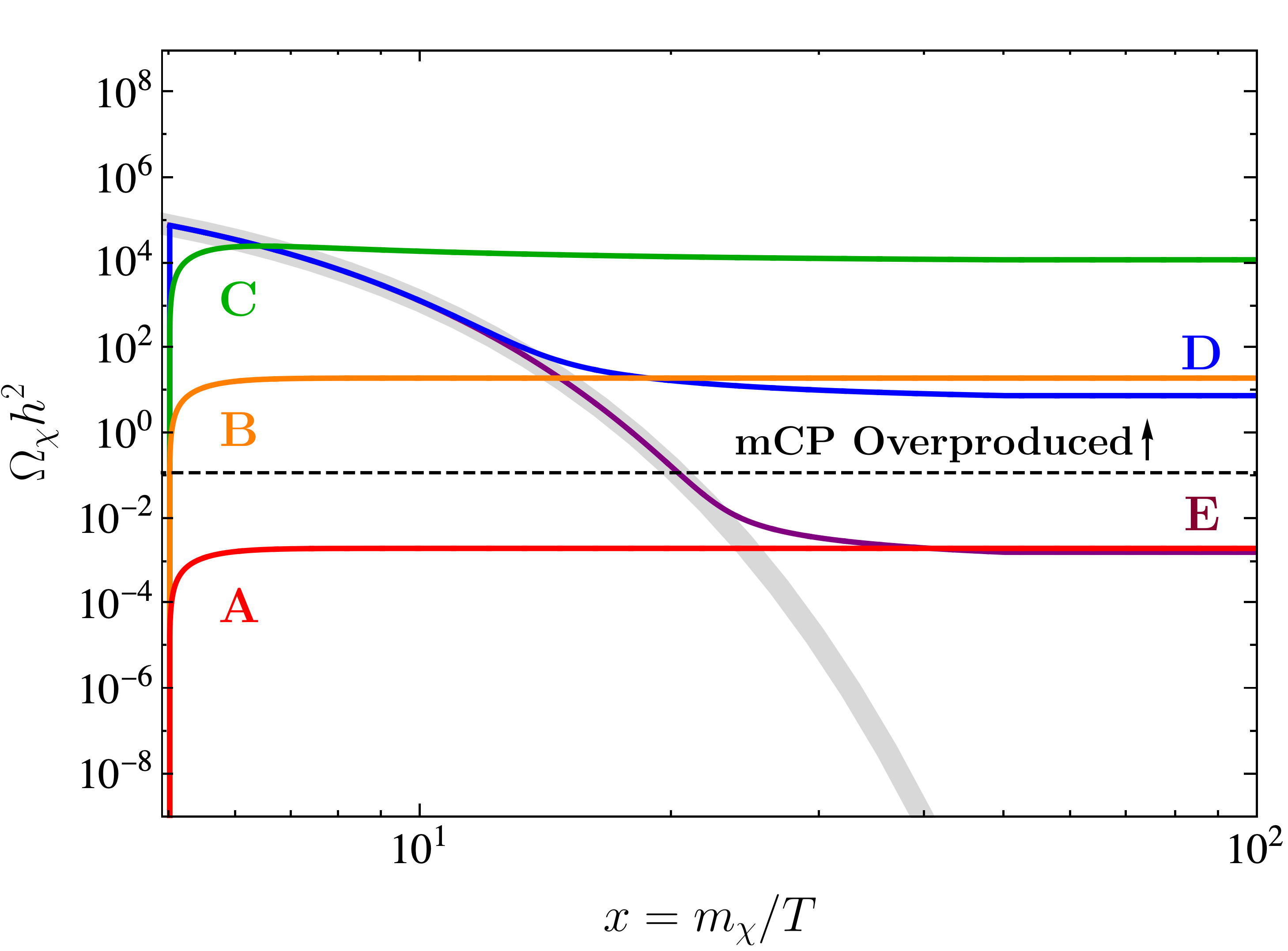}
\caption{ {\bf Left}: 
For $T_\rh = 10\, \MeV$, the contours of the ``pure'' mCP's late-time ($x\equiv m_\chi/T > 100$) relic abundances. The lower\,(upper) red curve is the freeze-in\,(freeze-out) curve. $\Omega_{\chi} h^2 \gg 0.12$ in the region enveloped by these two curves. 
The freeze-out and the freeze-in curves connect because larger $q_\chi$ is needed in compensation for the exponential suppression of the interaction rate when $m_\chi \gtrsim T_\rh$. 
The points B, C, and D are related to the overproduction of the mCPs, while A and E are related to the case in which mCPs are the subcomponent of the dark matter.
{\bf Right}: The cosmological evolutions of the abundances of ``pure'' mCPs start from zero abundances, thus the irreducible productions. 
Points A to E are all for $m_\chi=50\,\MeV$.
For curves A and B, mCP abundance quickly acquires the nonzero value at $T \simeq T_\rh$ and remains constant afterward, which corresponds to the UV freeze-in scenario, explained in the main text. D and E curves are related to the freeze-out scenario. Curve C is between the freeze-in and freeze-out scenarios.}
\label{fig:mcp_contour_Trh_10MeV}
\end{figure*}

For the mCP $\chi$, the Boltzmann equation describing its early universe evolution of the number density is
\bea
\label{eq:boltz_pure_mcp}
\dot{n}_\chi + 3 H n_\chi 
 & \simeq \mathcal{C}_n(T) \left( 1 - \frac{n_\chi^2}{n_{\chi,\eq}^2} \right),
\eea
where
\bea
\mathcal{C}_n(T) = 2   n_Z \langle \Gamma \rangle_{Z \rightarrow \chi \chibar} + 2  n_{f} n_{\fbar} \langle \sigma v \rangle_{f \fbar \rightarrow \chi \chibar} 
\eea
is the collision term of mCPs' number-changing processes in the forward direction, i.e., $\SM \rightarrow \text{Dark}$. In \Eq{boltz_pure_mcp}, $f=l,q,\nu$ are the SM fermions, and the factor ``2'' counts the production of $\chi$ and $\chibar$.  According to \cite{Gondolo:1990dk},  $\langle \sigma v \rangle$ is the thermally averaged cross-section, and it can be written as 
\bea
\label{eq:sigma_v}
\langle \sigma v \rangle = \frac{ \int_{4 m^2}^\infty ds \,\, \sigma(s) K_1\left(\frac{\sqrt{s}}{T}\right) \sqrt{s} ( s -4 m^2 ) }{ 8 T \left[ m^2 K_2\left(\frac{m}{T}\right) \right]^2 },
\eea
where $m=m_f$ for $f \fbar \rightarrow \chi \chibar$ channel. 
$\langle \Gamma \rangle$ is the thermally averaged decay rate, and it can be written as 
\bea
\langle \Gamma \rangle  = \frac{K_1\left(\frac{m}{T}\right)}{K_2\left(\frac{m}{T}\right)} \Gamma,
\eea

where $\Gamma = \Gamma_{Z \rightarrow \chi \chibar}$ for $Z \rightarrow \chi \chibar$ channel. As shown in \cite{Dvorkin:2019zdi}, in the mass region $m_\chi \gtrsim 1 \,\MeV$, the contribution from the on-shell plasmon decay $\gamma^* \rightarrow \chi \chibar$ is negligible; therefore, we don't consider this channel in our calculation. 
We solve the Boltzmann \Eq{boltz_pure_mcp} to determine the freeze-in and freeze-out curves in \Fig{bounds_Trh} to \ref{fig:mcp_contour_Trh_10MeV}.

Here, we can derive \Eq{YX_FI} and understand the dip below $m_Z/2$ for the freeze-in curves in \Fig{bounds_Trh} and \ref{fig:mcp_contour_high_Trh}. The mCP abundance in the freeze-in region can be written as
\bea
Y_\chi^{\text{FI}} \simeq \int^{T_\rh}_{T} dT \frac{1}{T H s} \mathcal{C}_n(T), \,\,\, T\gtrsim m_\chi,
\eea
neglecting the back-reaction and the variation of $g_*$ for analytical derivation.
Because $C_n(T) \sim q_\chi^2 \alphaem^2 T^4$, we have \Eq{YX_FI} when $T_\rh \gg m_\chi$. Because the annihilation channel $f \fbar \rightarrow \chi \chibar$ has the $Z$-resonance peak, $\langle \sigma v \rangle_{f \fbar \rightarrow \chi \chibar}$ is enhanced by a factor $m_Z^2/\Gamma_Z^2$, which explains the dip in the freeze-in curve when $m_\chi < m_Z/2$, as shown in the left panel of \Fig{mcp_contour_high_Trh}. Here, $\Gamma_Z$ and $m_Z$ are the $Z$ boson's decay width and mass, respectively. Because $Z$-resonance contribution to $\langle \sigma v \rangle_{f \fbar \rightarrow \chi \chibar}$ has an exponential suppression factor $K_1(m_Z/m_\chi) \sim \exp(-m_Z/m_\chi)$ as shown in \Eq{sigma_v}, the $Z$-resonance dip in the freeze-in curve disappears when $m_\chi \ll m_Z$. 

We also derive \Eq{mcp_yield_fo} and understand the dip at the $m_Z/2$ for the freeze-out curves in \Fig{bounds_Trh} and \ref{fig:mcp_contour_high_Trh}, by using the sudden freeze-out approximation. When $n_\chi \langle \sigma v \rangle_{f \fbar \rightarrow \chi \chibar} \sim H$, the mCPs' annihilation becomes inefficient and decoupled from the chemical equilibrium with the SM bath. At this time, $T \sim m_\chi$ and $\langle \sigma v \rangle_{f \fbar \rightarrow \chi \chibar} \sim q_\chi^2 \alphaem^2/ m_\chi^2$, therefore we can derive \Eq{mcp_yield_fo} given the hign $T_\rh$.  From the left panel of \Fig{mcp_contour_high_Trh}, we can also see that the freeze-out curve has a Z-dip with the width $\Delta m_\chi \sim \Gamma_Z$ and scaled by $\Gamma_Z/m_Z$ near $m_\chi \sim m_Z/2$.

In the final part of this section, we comment on the gravitational production of mCPs. For mCPs in the freeze-out region, the initial-condition dependence is washed out because the mCPs are in thermal equilibrium with the SM bath. For mCPs in the freeze-in region, the gravitational production of mCPs is highly suppressed with a low $T_\rh$~\cite{Kuzmin:1998kk,Kuzmin:1999zk,Chung:1998zb,Chung:2004nh,Chung:2011ck, Chung:2018ayg}, and is negligible in the scenarios we consider.

\newpage

\section{Energy injection to the dark radiation}
\label{appx:energy_inj}

For the mCP from kinetic mixing, the mCP and the dark photon can form a thermal bath when $g_d$ is large enough. To explore its $\Delta N_\eff$ bound, we calculate the energy transfer from the SM bath to the dark bath following the discussion in  \cite{Vogel:2013raa, Kuflik:2015isi, Kuflik:2017iqs, Luo:2020fdt, Luo:2020sho, Fernandez:2021iti, Adshead:2022ovo}. The Boltzmann equation describing the energy transfer from the SM bath to the dark bath is
\bea
\label{eq:energy_trans_dark_bath}
&\dot{\rho}_{d} + 3 (\rho_d + p_d) H \rho_{d} \simeq \mathcal{C}_\rho(T) - \mathcal{C}_\rho(T_d)
\eea
where
\bea
\mathcal{C}_\rho = n_Z \langle \Gamma E \rangle_{Z \rightarrow \chi \chibar} + n_{f} n_{\fbar} \langle \sigma v E \rangle_{f \fbar \rightarrow \chi \chibar}
\eea
is the collision term of the energy transfer in the forward direction. In \Eq{energy_trans_dark_bath}, $p_d \simeq \rho_d/3$ for the dark radiation, and $H^2 \simeq (8\pi/3\mpl^2)(\rho_\rad + \rho_d) $. 
From \cite{Gondolo:1990dk,Chu:2011be}, we have
\bea
\langle \sigma v E \rangle = \frac{ \int_{4 m^2}^\infty ds \,\, \sigma(s) K_2\left(\frac{m}{T}\right) s (s-4m^2) }{ 8 T \left[ m^2 K_2\left(\frac{m}{T}\right) \right]^2 }
\eea
for the energy transfer through $f \fbar \rightarrow \chi \chibar$ channel, and
\bea
\langle \Gamma E \rangle = m \Gamma
\eea
for the energy transfer through $Z \rightarrow \chi \chibar$. Because we focus on the boundary of the $\Delta N_\eff$ bound, $\rho_d \ll \rho_\rad$, therefore dark radiation's contributions to the Hubble parameter and the total entropy are negligible. Given this, for $m_\chi \lesssim 1\,\GeV$, the dark photon energy density can be written in the integrated form
\bea
\label{eq:int_rho_d}
\frac{\rho_{d}}{s^{4/3}} & \simeq \int^{T_\rh}_{T} \frac{dT}{T H s^{4/3}} \mathcal{C}_\rho(T), \,\,\, T\gtrsim m_\chi,
\eea
once $T \sim T_d$, the dark bath and the SM bath will reach the thermal equilibrium.

Using the definition~\cite{Boehm:2012gr}
\bea
\label{eq:DelNeff_Def}
\Delta N_\eff = \frac{\rho_{d}}{\rho_\nu}, \quad \text{where $\rho_\nu = \frac{7}{8} \rho_\gamma \left(\frac{T_\nu}{T}\right)^4$},
\eea
$\mathcal{C}_E(T) \sim q_\chi^2 \alphaem^2 T^5$, and order one $g_{*,S}$ variation , we can derive the $\Delta N_\eff$ bound for the mCP with kinetic mixing as
\bea
\Delta N_\eff \sim q_\chi^2 \alphaem^2 \frac{\mpl}{m_\chi}
\eea
in the region $\Delta N_\eff \lesssim 1$ when $m_\chi \lesssim T_\rh$.

\clearpage

\bibliography{references}

\begin{thebibliography}{100}
\bibitem{UV_freezein}
In this context, the UV freeze-in refers to the situation where most of the BSM
  particles are produced at $T\sim T_\rh$, compared with the (IR) freeze-in
  where most of the particles are produced at $T \sim m_\chi$.
\end{thebibliography}

\end{document}